\documentclass[useAMS,usegraphicx,usenatbib]{mn2e}          % ,referee

\usepackage{times}

\newcommand{\Msun}{\ensuremath{\,{\rm M}_\odot}}                  % Solar mass symbol
\newcommand{\Rsun}{\ensuremath{\,{\rm R}_\odot}}                  % Solar radius symbol
\newcommand{\Teff}{\ensuremath{T_{\rm eff}}}                      % Effective temperature symbol
\newcommand{\Mjup}{\ensuremath{\,{\rm M}_{\rm Jup}}}              % Jupiter mass symbol
\newcommand{\Rjup}{\ensuremath{\,{\rm R}_{\rm Jup}}}              % Jupiter radius symbol
\newcommand{\kms}{\,km\,s$^{-1}$}                                 % km/s symbol
\newcommand{\ms}{\,m\,s$^{-1}$}                                   % m/s^2 symbol
\newcommand{\mss}{\,m\,s$^{-2}$}                                  % m/s^2 symbol
\newcommand{\as}{\ensuremath{^{\prime\prime}}}                    % Arcsecond symbol
\newcommand{\am}{\ensuremath{^\prime}}                            % Arcminute symbol
\newcommand{\FeH}{\ensuremath{\left[\frac{\rm Fe}{\rm H}\right]}} % [Fe/H] symbol
\newcommand{\pjup}{\ensuremath{\,\rho_{\rm Jup}}}                 % Jupiter density symbol
\newcommand{\psun}{\ensuremath{\,\rho_\odot}}                     % Solar density symbol
\newcommand{\mcc}[1]{\multicolumn{3}{c}{#1}}

\newcommand{\ermcc}[5]{\mcc{\ensuremath{{#1\,^{+#2}_{-#3}}\,^{+#4}_{-#5}}}}

\title[Transits and starspots of WASP-19]
      {Transits and starspots in the WASP-19 planetary system}

\author[Tregloan-Reed et al.]
       {Jeremy Tregloan-Reed\,$^{1}$\thanks{Email: j.j.tregloan-reed@keele.ac.uk}, John Southworth\,$^{1}$, C. Tappert\,$^{2}$ \\
        $^{1}$\,Astrophysics Group, Keele University, Staffordshire, ST5 5BG, UK \\
        $^{2}$\,Departamento de F\'{i}sica y Astronom\'{i}a, Universidad de Valpara\'{i}so, Avda. Gran Breta\~na 1111, Valpara\'{i}so, Chile \\
}

\begin{document} \maketitle %%%%%%%%%%%%%%%%%%%%%%%%%%%%%%%%%%%%%%%%%%%%%%%%%%%%%%%%%%%%%%%%%%%%%%%%%%%%%%%%%%%%%%%%%%%%%%%%%%%%%%%%%%%%%%%%%%%%%%%%%
%%%%%%%%%%%%%%%%%%%%%%%%%%%%%%%%%%%%%%%%%%%%%%%%%%%%%%%%%%%%%%%%%%%%%%%%%%%%%%%%%%%%%%%%%%%%%%%%%%%%%%%%%%%%%%%%%%%%%%%%%%%%%%%%%%%%%%%%%%%%%%%%%%%%%

\begin{abstract}
We have developed a new model for analysing light curves of planetary transits when there are starspots on the stellar disc. Because the parameter space contains a profusion of local minima we developed a new optimisation algorithm which combines the global minimisation power of a genetic algorithm and the Bayesian statistical analysis of the Markov chain. With these tools we modelled three transit light curves of WASP-19. Two light curves were obtained on consecutive nights and contain anomalies which we confirm as being due to the same spot. Using these data we measure the star's rotation period and velocity to be $11.76 \pm 0.09$\,d and $3.88 \pm 0.15$\kms, respectively, at a latitude of 65$^\circ$. We find that the sky-projected angle between the stellar spin axis and the planetary orbital axis is $\lambda = 1.0^{\circ} \pm 1.2^{\circ}$, indicating axial alignment. Our results are consistent with and more precise than published spectroscopic measurements of the Rossiter-McLaughlin effect. 
\end{abstract}


%%%%%%%%%%%%%%%%%%%%%%%%%%%%%%%%%%%%%%%%%%%%%%%%%%%%%%%%%%%%%%%%%%%%%%%%%%%%%%%%%%%%%%%%%%%%%%%%%%%%%%%%%%%%%%%%%%%%%%%%%%%%%%%%%%%%%%%%%%%%%%%%%%%%%

%%%%%%%%%%%%%%%%%%%%%%%%%%%%%%%%%%%%%%%%%%%%%%%%%%%%%%%%%%%%%%%%%%%%%%%%%%%%%%%%%%%%%%%%%%%%%%%%%%%%%%%%%%%%%%%%%%%%%%%%%%%%%%%%%%%%%%%%%%%%%%%%%%%%%

\section{Introduction}                                                                                                              
\label{sec:intro}

It was \citet{Silva2003} who first put forward the idea of using planetary transits to detect starspots, through an anomalous brightening when the planet passes over the starspot during the transit of the host star. As transit surveys such as WASP \citep{SuperWasp} and HAT \citep{HAT} detect ever more transiting planets, the number of known planets orbiting an active star will increase. \citet{Rabus2009,Pont2007} and \citet{Winn2010b} have all shown how a planet crossing a starspot during transit can create a small increase in the received flux from the star. This occurs because the spot is generally cooler than the surrounding photosphere, so less light is lost when a planet is occulting the spot than when the planet is in front of the unspotted parts of the photosphere. One of the remaining problems is to model the effects of both the transit and spot accurately and precisely.

\citet{Sanchis2011a}, \citet{Nutzmann2011} and \citet{Desert2011} used photometric observations to show that it is possible to calculate the obliquity of the system when there are light curves of multiple transits affected by the same spot(s). A large spin-orbit misalignment has been found in this way for HAT-P-11 \citep{Sanchis2011b}. For a multiple planetary system, starspots have also been used to test the alignment of the stellar spin axis against the orbital planes of the planets \citep{Sanchis2012}. The obliquity of a planetary system helps understand which mechanism was predominant in the dynamical evolution of the system \citep{Sanchis2012,Winn2010c}. A low obliquity would follow from the idea that the planet formed at a large distance from its host star and, through tidal interactions with the protoplanetary disc, suffered orbital decay. Larger obliquities are expected when orbital decay occurred due to gravitational interactions from other bodies in the system \citep{Schlaufman2010}.

At present there are three main ways to measure the rotation period of a star. The first is to use photometric rotational modulation over many months or years \citep{Hall1972}. The second method uses radial velocity mesurements to find the projected rotational velocity, $v\,\sin I$, which gives a lower limit on $v$ and thus the rotation period. The third method, presented by \citet{Silva2008}, is the idea of measuring the rotation period of a star by using a transiting planet crossing a starspot. This opens up the possibility to allow the rotation period of a star to be found from two sets of photometry from a 2m-class ground-based telescope. 

Once the rotation period of a star is known it is possible to find the age of the star according to the gyrochronology relationship \citep[e.g.]{Barnes2007}. It is also possible, when combined with $v\,\sin I$, to find $I$, the inclination of the stellar spin axis towards the observer. To find the spin-orbit alignment of a system, \citet{Fabrycky2009} showed that we require three parameters, orbital inclination $i$, the inclination of the stellar spin axis towards the observer $I$, and the sky-projected spin-orbit alignment $\lambda$. The quantity $\lambda$ can be found by using two different methods. The first method is the Rossiter-McLaughlin effect \citep{RMa,RMb}, while the second method uses planetary transits crossing over starspots \citep{Sanchis2011a,Sanchis2011b,Sanchis2012,Nutzmann2011,Desert2011}.

After observing three light curves of WASP-19 with the aim of obtaining accurate physical properties, we discovered that two of our datasets contained a starspot anomaly. Starspots can affect the shape of a transit \citep{Silva2010} and if not correctly modelled can lead to biased measurements of the system parameters. To achieve our original goal to obtain precise measurements of the system properties we decided to develop a new model capable of modelling both the transit and starspots simultaneously. With a precise known position of the spot at two close but distinct times we would then be able to calculate the obliquity of the system and compare this to the values found from measurement of the Rossiter-McLaughlin effect \citep{Hellier2011,Albrecht2012} and check if WASP-19 follows the theory put forward by \citet{Winn2010c} that cool stars will have low obliquity (see also \citealt{Triaud11aa}). This would also allow us to measure the rotation period of the star and compare it to the value found by photometric modulation \citep{Hebb2010}.

Section\,\ref{sec:prism} describes the \textsc{idl}\footnote{The acronym {\sc idl} stands for Interactive Data Language and is a trademark of Exelis Visual Information Solutions. For further details see {\tt http://www.ittvis.com/ProductServices/IDL.aspx}.} code \textsc{prism} and how it models both the planetary transit and the star-spot. Section\,\ref{sec:fitting} describes the optimisation algorithm used to find the optimal solution together with their associated uncertainties. Section\,\ref{sec:data} gives an overview of the observations and the manner in which they were collected. Section\,\ref{sec:results} shows the best fitting results of the model for both the system and starspots. Section\,\ref{sec:Conclusions} reviews previous work on WASP-19 and compares the results of this work.

%%%%%%%%%%%%%%%%%%%%%%%%%%%%%%%%%%%%%%%%%%%%%%%%%%%%%%%%%%%%%%%%%%%%%%%%%%%%%%%%%%%%%%%%%%%%%%%%%%%%%%%%%%%%%%%%%%%%%%%%%%%%%%%%%%%%%%%%%%%%%%%%%%%%%

\section{Modelling transits and starspots: introducing \textsc{prism}}
\label{sec:prism}

\begin{figure} \includegraphics[width=0.46\textwidth,angle=0]{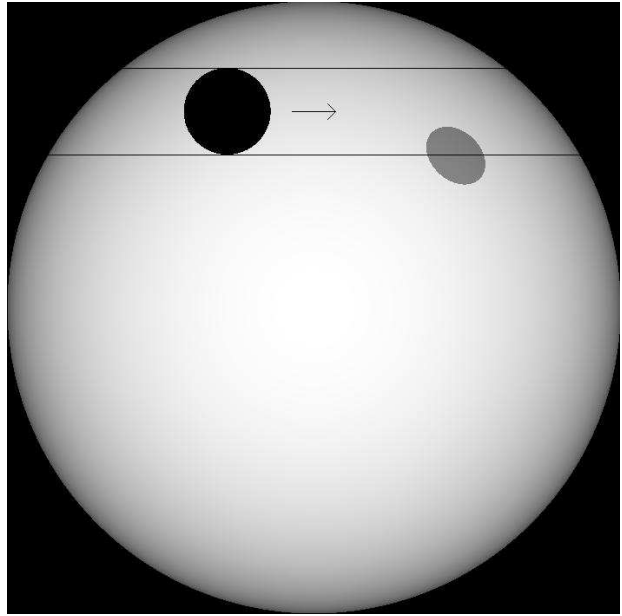}
\caption{\label{fig:prism-out} An output model of a transit coupled with a starspot 
using \textsc{prism}. The transit chord is represented by the two horizontal black 
lines. The black disc to the left is the planet. A dark starspot has been placed 
on the curved stellar surface to show how \textsc{prism} projects an elliptical 
shape onto the stellar disc for a circular spot.} \end{figure}

When dealing with starspot anomalies in transit data a common course of action is to model the transit first and then deal with the starspot based on the residuals \citep[e.g.][]{Sanchis2011a,Maciejewski2011}. This can unfortunately lead to unknown uncertainties and biases in the measurements of the stellar and planetary radii, inclination and limb darkening coefficients. \citep[e.g.][]{Ballerini2012}. This is because when a starspot is on the visible part of the stellar disc it reduces the received flux by an amount $\Delta F_{\rm spot}$. When the planet transits the star it blocks $\Delta F_{\rm planet}$ of the stellar flux. The depth of the transit is the fractional amount of flux blocked by the planet, which depends on the ratio of the areas of the planet and star. Without a starspot and in the absence of limb darkening (LD) the equation for the ratio of the radii is:
\begin{equation} 
\left(\frac{R_{\rm p}}{R_{\rm s}}\right)^{2} = \frac{\Delta F_{\rm planet}}{F}
\end{equation}
where $F$ is the total flux of the unspotted star and $R_{\rm p}$ and $R_{\rm s}$ are the radii of the planet and star. If a starspot is placed on the stellar disc and causes a decrease in stellar flux of $\Delta F_{\rm spot}$, the above equation becomes:
\begin{equation}
\alpha \left(\frac{R_{\rm p}}{R_{\rm s}}\right)^{2} = \frac{\Delta F_{\rm planet}}{\left(F - \Delta F_{\rm spot}\right)}
\end{equation}
where $\alpha$ is the ratio of the transit depths in the spotted and unspotted cases. Because $\Delta F_{\rm spot} > 0$ for a cool spot, the transit gets deeper ($\alpha > 1$). Neglecting this would result in an incorrect measurement of the ratio of the radii, $\frac{R_{\rm p}}{R_{\rm s}}$.

To obtain accurate measurements of the system and spot parameters we created an \textsc{idl} computer code to model both the planetary transit and starspots on the stellar surface. \textsc{prism}\footnote{Available from http://www.astro.keele.ac.uk/$\sim$jtr} (Planetary Retrospective Integrated Star-spot Model) uses a pixellation approach to create the modelled star on a two-dimensional array in Cartesian coordinates (see Fig.\,\ref{fig:prism-out}). This makes it possible to model the transit, limb darkening and starspots on the stellar disc simultaneously. \citet{Silva2003} used a similar model to describe the starspots on HD\,209458, but with the drawback of having to use fixed system parameters from previous results. \textsc{prism} is set to use the standard quadratic limb darkening law and uses the fractional stellar and planetary radii (the radii scaled by the semimajor axis, $r_{\rm s,p} = R_{\rm s,p}/a$). \textsc{prism} requires ten parameters to model the system:-- 
\begin{itemize}
\item the ratio of the radii, $\frac{r_{\rm p}}{r_{\rm s}} = \frac{R_{\rm p}}{R_{\rm s}}$
\item the sum of the fractional radii, $r_{\rm p} + r_{\rm s} = \frac{R_{\rm p}+R_{\rm s}}{a}$
\item the linear limb darkening coefficient, $u_1$
\item the quadratic limb darkening coefficient, $u_2$
\item the orbital inclination, $i$
\item a reference transit midpoint, $T_{0}$
\item the longitude of the centre of the spot, $\theta$. Longitude is defined to be zero degrees at the centre of the stellar disc.
\item the latitude of the centre of the spot, $\phi$. Latitude is defined to be zero degrees at the north pole and 180 degrees at the south pole.
\item the spot size, $r_{\rm spot}$, in degrees.
\item the spot contrast, $\rho_{\rm spot}$, which is the surface brightness of the spot versus the immaculate photosphere
\end{itemize}
To ensure sufficient numerical resolution, the diameter of the planet is hard-coded to be 100 pixels, and the size of the star in pixels is scaled according to the specified ratio of the radii. When modelling a starspot, \textsc{prism} projects a circular spot onto the curved surface of a star. Because of this it is able to account for spots which are visible on the edge of the star even if their centre is beyond the limb.

\subsection{Sample light curves}

\begin{figure} \includegraphics[width=0.46\textwidth,angle=0]{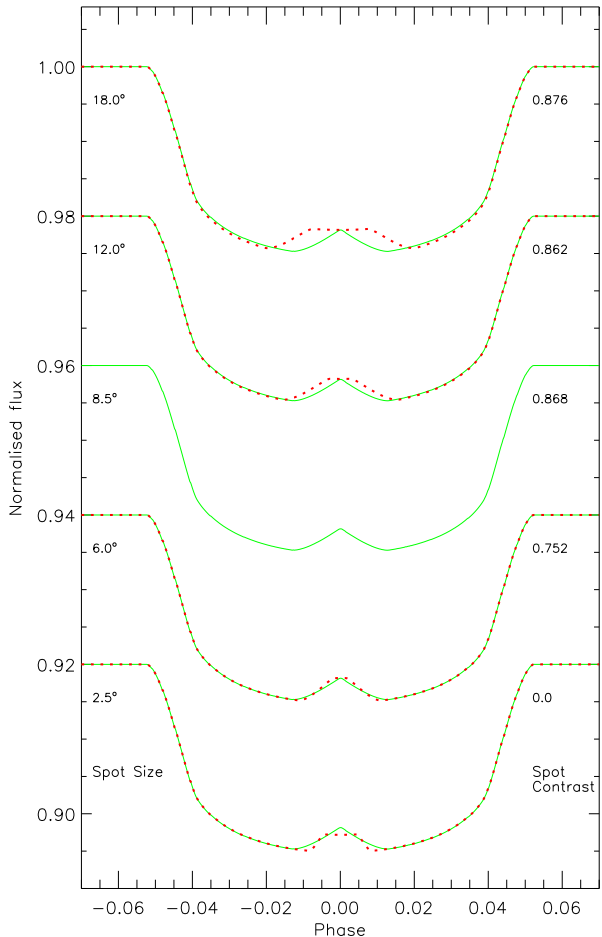}
\caption{\label{fig:spot-curves} Five light curves showing how the shape of the spot anomaly 
changes with the size of the spot relative to that of the planet. The spot contrast was also 
modified for each light curve to maintain an approximately constant spot anomaly amplitude. 
This amplitude gives a lower limit for the size of the spot. The spot sizes are labelled on 
the left of the plot and the spot constrasts on the right. The $8.5^{\circ}$ spot (green solid 
line) is the representation of when the spot is of equal size to the transiting planet. There 
is a degeneracy between the spot radius and contrast, which can be broken when modelling data 
of sufficiently high precision and cadence.} \end{figure}

When a spot anomaly is viewed during a transit the total flux received increases for a dark spot. The total change in flux is based on the surface area and contrast of the spot \citep{Silva2003}. Therefore there is a degeneracy between these two parameters. Fig.\,\ref{fig:spot-curves} shows example light curves for anomalies of approximately the same amplitude and due to a range of spot sizes and constrast. It is possible to discern three regimes from this diagram. Firstly, when the spot is of a similar size to the planet the shape of the spot-occultation is an inverted `V'. This is due to the fact that the amount of time the planet spends fully eclipsing the spot is very small compared to the duration of the partial eclipse phases. Secondly, for a larger spot, both the peak and base of the spot-transit increase, because the planet reaches the spot earlier and spends more time fully eclipsing the spot. Thirdly, for a smaller spot, the peak broadens due to the planet fully eclipsing the spot for longer while the base shortens due to the fact that the total duration is shorter. These three distinct shapes allow the degeneracy between the spot radius and contrast to be broken for data of sufficient precision and time sampling.

It is also apparent that the amplitude of the spot-transit gives a lower limit on the size of the spot, below which the spot is too small to give such an amplitude even if its contrast is zero. In Fig.\,\ref{fig:spot-curves} the $2.5^{\circ}$ spot has a contrast of zero and is still unable to achieve the same change in flux as the other spots.

%%%%%%%%%%%%%%%%%%%%%%%%%%%%%%%%%%%%%%%%%%%%%%%%%%%%%%%%%%%%%%%%%%%%%%%%%%%%%%%%%%%%%%%%%%%%%%%%%%%%%%%%%%%%%%%%%%%%%%%%%%%%%%%%%%%%%%%%%%%%%%%%%%%%%

\section{Optimisation algorthms: introducing \textsc{gemc}}
\label{sec:fitting}

Our first attempt at fitting real data with {\sc prism} utilised a Monte Carlo Markov Chain (MCMC) algorithm. This was introduced in order to use Bayesian methods to find the best fit and associated errorbars. We found that the problem with this approach was that the many local minima in the parameter space tended to trap our MCMC chains, resulting in poor mixing and convergence. This could be solved by using a large number of iterations, but such an approach was ill-suited to {\sc prism} due to the significant amount of processing time required per iteration\footnote{A single evaluation of a model appropriate for WASP-19, with 70 datapoints, takes {\sc prism} typically 0.7\,s using a 2.7\,GHz duel core desktop computer}. We found that MCMC chains required up to $10^6$ iterations to converge properly, depending on how often they got stuck in local minima, which equated to about a week of calculation time.

Our solution to this problem was to implement a genetic algorithm (GA). A GA mimics biological processes by spawning successive generations of solutions based on breeding and mutation operators from the previous generation. By performing these operations the new solutions are generated based on the fitness of the parent solutions, not a perturbation of their parameters. Because of this a GA can be considered as a global optimiser where solutions can jump large distances across the solution space. The efficiency of a GA at finding the global solution is demonstrated by \citet{Char1995} but, as discussed by \citet{Rajpaul2012}, it does have some limitations. These are primarily that GAs are ill-suited to Bayesian statistics, and that they are good at finding where the global solution is but poor at locating its exact position. 

Our initial answer to the latter problem was to allow the GA to find and constrain the global solution and then to use the MCMC algorithm to perform the error analysis for this solution. This allowed us to reduce the computation time from seven to five days. Dissatisfied with the fact that two different optimisation algorithms had to be used, one to locate the global solution and the other to obtain parameter uncertainties, we decided to develop a new optimisation algorithm, which combined the global optimisation power of the GA but also able to perform Baysian statistics on the solutions. We call this new algorithm \textsc{gemc}\footnote{Available from http://www.astro.keele.ac.uk/$\sim$jtr} (Genetic Evolution Markov Chain). \textsc{gemc} is based on a Differential Evolution Markov Chain (DE-MC) put forward by \citet{Cajo2006}.

\textsc{gemc} begins by randomly generating parameters for $N$ chains, within the user-defined parameter space, and then simultaneously evolves the chains for $X$ generations. At each generation the chains are evaluated for their fitness\footnote{A solution's fitness was found by calculating the $1 / \chi^2$ value.}. The parameters of the fittest member undergo a $\pm 1\%$ perturbation and its fitness is then re-evaluated. If the fitness has improved it is accepted but if the fitness has deteriorated it is accepted based on a Gaussian probability distribution:
\begin{equation}
P = \exp\left(\frac{\left(\chi^2_{(n-1)} - \chi^2_{(n)}\right)}{2}\right)
\end{equation}
where $(n-1)$ is the previous generational chain and $n$ is the current generational chain being evaluated. The next step is to then evolve the other chains. This is accomplished in a similar way as a GA, in that the chain parameters are modified by incorporating the parameters of another chain. But unlike a GA where a member is picked by a weighted random number and then the digits of each parameter are crossed over with the digits from a different member, \textsc{gemc} directly perturbs the parameters of each chain in a vector towards the best-fitting chain. The size of this perturbation is between zero and twice the distance to the best-fitting chain, allowing the chain to not only move towards but also to overshoot the position of the best-fitting chain. 

An example would be a two-dimensional function $f(x,y)$. The difference between a given chain and the best-fitting chain in this case is $(\Delta x,\Delta y)$. This difference is then multiplied by a random scalar $\gamma$ for each parameter, where $\gamma$ is in the interval [0,2], and then added to the given chain's parameters $(x_0,y_0)$ to form the new potential solution $f(x_1,y_1)$.
\begin{equation}
x_{1} = x_{0} + \gamma_{x} \Delta x
\end{equation}
\begin{equation}
y_{1} = y_{0} + \gamma_{y} \Delta y
\end{equation}
When $\gamma = 0$ the parameter is not perturbed, $\gamma = 1$ the parameter equals the current best fitting value and when $\gamma = 2$ the parameter is perturbed to the opposite position. This allows the potential solution to travel large distances across the parameter space unimpeded by local peaks. After the parameters have been perturbed the chain is then re-evaluated and is selected using the same method as the best fitting chain.

\textsc{gemc} runs in two stages. The first stage, called the `burn in', is used to find the optimal solution to the data using the above method. After this the second stage starts in which each chain undertakes an independent MCMC run. The starting points for each MCMC chain lie close but not exactly at the optimal solution. In essence what we have is the same outcome from running a GA to find the best fitting solution and to use this to tightly constrain the starting parameter range of an MCMC run.

When \textsc{gemc} was used in conjunction with \textsc{prism} to find the best fitting solution to the same dataset as above, the computational time reduced dramatically, from five days to 14 hours using a large parameter range (see Section\,5). When the parameter range was set to the same as used by the GA or the MCMC, \textsc{gemc} was able to produce the best fitting solution and similar uncertainties in the fitted parameters as the MCMC within 10 hours\footnote{\textsc{gemc} is able to produce similar parameter uncertainties as an MCMC run with only 1000 iterations (taking 11\,min to calculate), but for statistical certainty the MCMC section of \textsc{gemc} was allowed to run for 50\,000 iterations (9.3\,hr).}.

\begin{figure} \includegraphics[width=0.46\textwidth,angle=0]{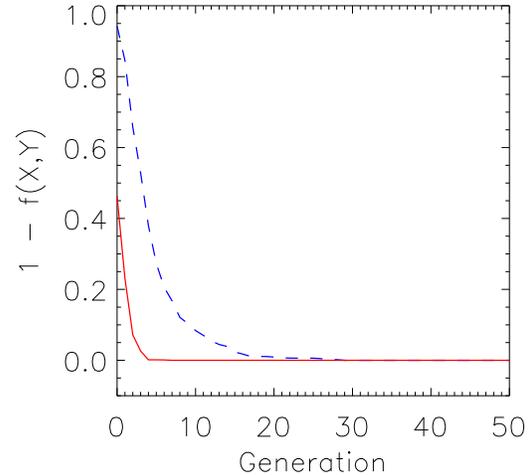}
\caption{\label{fig:evolution} The evolution of the fittest chain (solid line) 
and the mean fitness chain (dashed line) from each generation. The maximum peak 
was found in five generations. The fitness is measured as $1 - f(x,y)$.} \end{figure}

\begin{figure*} \includegraphics[width=0.92\textwidth,angle=0]{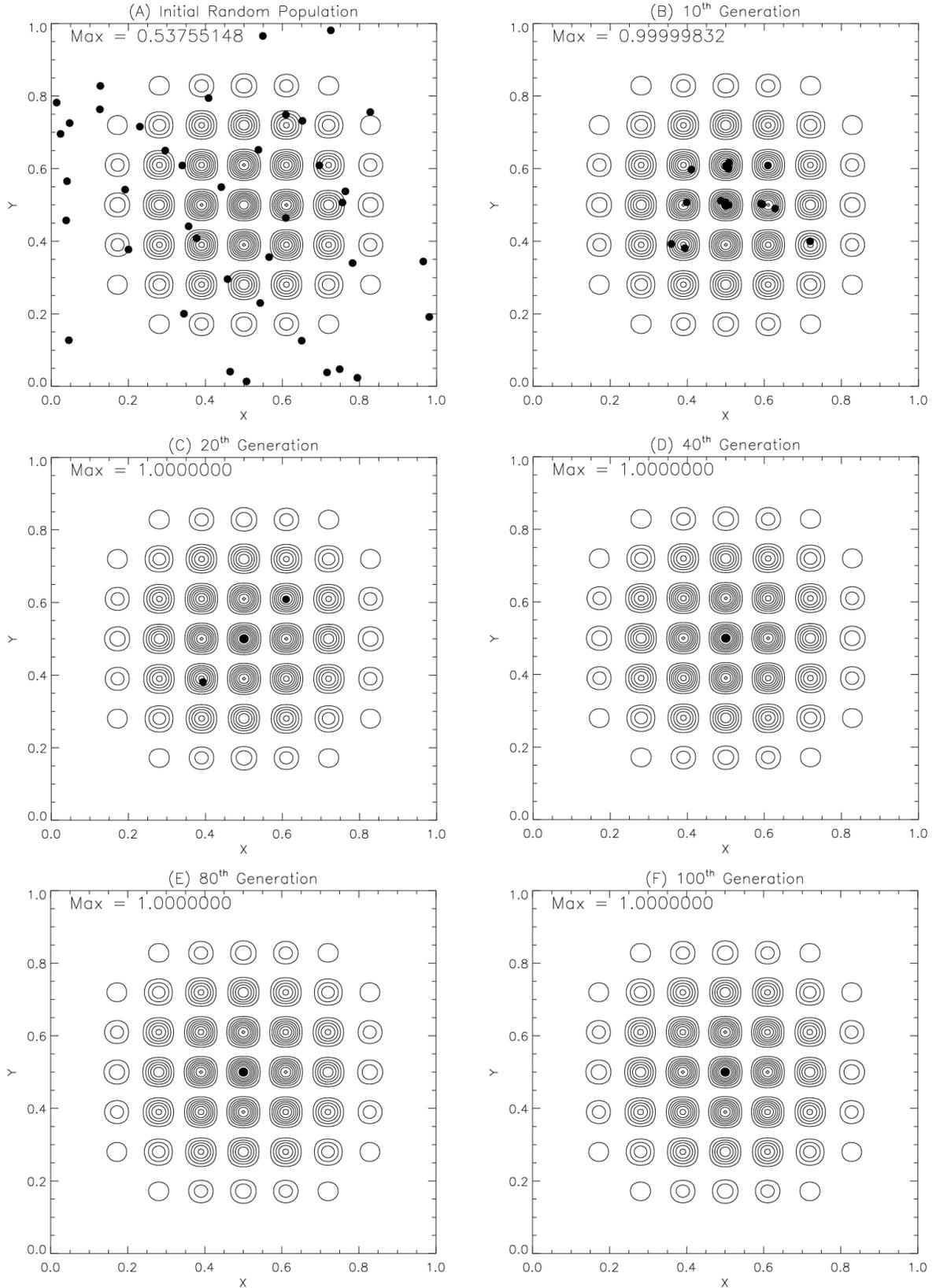}
\caption{\label{fig:pikaia-results} For a comparison with \citet{Char1995}. {\sc gemc} results for $N=40$ chains and for 
$X = 100$ generations. The global maximum peak and global maximum point have 
been discovered by the 10th and 20th generations respectively. By the 40th 
generation all 40 chains have found the global maximum point.} \end{figure*}

To demonstrate the power of \textsc{gemc} at finding the optimal solution of a rugged parameter space we chose to test it against the function used to test the genetic algorithm \textsc{pikaia} by \citet{Char1995};
\begin{eqnarray}
f(x,y) = [16x(1-x)y(1-y)\sin(n\pi x)\sin(n\pi y)]^2 \nonumber \\
x,y \in [0,1], \ n=1,2,...
\end{eqnarray}
The optimal solution to this function lies at the centre ($f(0.5,0.5) = 1$). \citet{Char1995} showed that it took \textsc{pikaia} with a population of 100 solutions up to 20 generations to find the global maximum peak but even after 100 generations it still had not found the global maximum point, confirming the GA inability to find best solutions with precision. While looking at Fig.\,\ref{fig:pikaia-results} we can clearly see that \textsc{gemc}, using a population of only 40 solutions, has found the global maximum peak within 10 generations and then went on to find the global maximum point within 20 generations. We can also see from Fig.\,\ref{fig:evolution} the power of \textsc{gemc}. The global maximum peak was actually found at the fifth generation and all solutions were very close to the global maximum point by the twentieth generation. This performance indicates that the required burn-in for \textsc{gemc} is extremely short and as such greatly reduces the computing time required to find the global solution.

%\begin{figure} \includegraphics[width=\columnwidth,angle=0]{pikaia.eps}
%\caption{\label{fig:pikaia} The function used to test the optimisation power 
%of \textsc{gemc}. Here $n=9$ to generate 81 local maxima.} \end{figure}

%%%%%%%%%%%%%%%%%%%%%%%%%%%%%%%%%%%%%%%%%%%%%%%%%%%%%%%%%%%%%%%%%%%%%%%%%%%%%%%%%%%%%%%%%%%%%%%%%%%%%%%%%%%%%%%%%

\section{Observations and data reduction}
\label{sec:data}

\begin{table*} \centering
\caption{\label{tab:obslog} Log of the observations presented in this work. $N_{\rm obs}$ is
the number of observations. `Moon illum.' and 'Moon dist.' are the fractional illumination of
the Moon, and its distance from WASP-19 in degrees, at the midpoint of the transit.}
\begin{tabular}{lcccccccccc} \hline
Date & Start time & End time &$N_{\rm obs}$& Exposure & Filter & Airmass & Moon & Moon & Aperture   & Scatter \\
     &    (UT)    &   (UT)   &             & time (s) &        &         &illum.& dist.& sizes (px) & (mmag)  \\
\hline
2010/02/24 & 06:18 & 09:34 &  68 & 90     & Gunn $r$ & 1.14 $\to$ 2.30 & 0.742 & 85.5 & 42, 60, 100 & 0.573 \\
2010/02/25 & 00:44 & 04:26 &  76 & 60--90 & Gunn $r$ & 1.40 $\to$ 1.04 & 0.818 & 78.1 & 52, 70, 90  & 0.464 \\
2010/02/28 & 04:01 & 07:41 &  74 & 90     & Gunn $r$ & 1.04 $\to$ 1.42 & 0.996 & 53.0 & 44, 64, 88  & 0.499 \\
\hline \end{tabular} \end{table*}

Three transits of WASP-19 were observed in February 2010 using the 3.6\,m New Technology Telescope (NTT) operated at ESO La Silla, Chile. The instrument used was EFOSC2, operated in imaging mode and with a Gunn $r$ filter (ESO filter \#786). In this setup the CCD covers a field of view of $(4.1\am)^{2}$ with a plate scale of 0.12\as\,px$^{-1}$. No binning or windowing was used, resulting in a dead time between consecutive images of 83\,s. The exposure time duration were 60-90\,s. The moon was bright and relatively close to the target star. The pointing of the telescope was adjusted to allow five good comparison stars to be observed simultaneously with WASP-19 itself. We were able to keep the telescope autoguiding through all observations. An observing log is given in Table\,\ref{tab:obslog}.

These observations were experimental for two reasons. Firstly, the NTT is an alt-az telescope fitted with an image derotator. This means that the path of light from each star through the telescope is continually changing, raising the possibility of correlated noise due to any optical imperfections. Secondly, the NTT is fitted with an actively controlled thin primary mirror designed to provide the best possible focus for normal observing strategies. Defocussing such a telescope might lead to a point spread function (PSF) which is variable in time, and thus correlated noise via flat-fielding errors.

In practise we found that, whilst careful attention had to be paid to the amount of defocussing, the NTT is perfectly capable of producing high-quality light curves whilst a long way out of focus due to stable symmetric PSFs. Our observations used this approach and are not plagued by correlated noise. This situation is similar to that of \citet{Winn2009}, who successfully observed WASP-4 using the Magellan Baade telescope. By contrast, \citet{Gillon2009} encountered serious problems in obtaining photometry of WASP-4 and WASP-5 with the ESO Very Large Telescope. This problem was attributed to the need to turn off the active optics system in order to achieve strong defocussing, and our results support the contention that this is not a general problem with alt-az telescopes or active-optics systems.

We reduced the data in an identical fashion to \citet{Sou2009,Sou2009b,Sou2009c,Sou2010b}. In short, we performed aperture photometry using an {\sc idl} implementation of {\sc daophot} \citep{Stetson1987}, and adjusted the aperture sizes to obtain the best results (see Table\,\ref{tab:obslog}). A first order polynomial was then fitted to the outside-transit data whilst simultaneously optimising the weights of the comparison stars. The resulting data have scatters ranging from 0.464 to 0.573 mmag per point versus a transit fit using \textsc{prism}. The timescale used is HJD/UTC.

%%%%%%%%%%%%%%%%%%%%%%%%%%%%%%%%%%%%%%%%%%%%%%%%%%%%%%%%%%%%%%%%%%%%%%%%%%%%%%%%%%%%%%%%%%%%%%%%%%%%%%%%%%%%%%%%%

\section{Data analysis}
\label{sec:results}

We began by selecting a search space for each parameter. As discussed in Section\,\ref{sec:fitting}, the ability of \textsc{gemc} to find the global minima in a short amount of computing time meant we were able to search a large area of parameter space to avoid the possibility of missing the best solution. The parameter search range used in analysing the WASP-19 datasets are given in Table\,\ref{tab:results}. 

First we modelled the three datasets of WASP-19 separately using \textsc{prism}, finding that the modelled parameters were within 1-$\sigma$ of each other (Table\,\ref{tab:results}). We then modelled all three datasets simultaneously. The ensuing parameters agreed with the individual results found previously, but we were unable to get as good a fit to the data. The reason for this seems to be the LD coefficients, which are in comparatively poor agreement when the three light curves are fitted individually. The scatter around the weighted mean is $\chi^2_\nu = 2.2$ for the linear coefficient and $1.9$ for the quadratic coefficient. This situation could be caused by the influence of the starspot on the LD coefficients. \citet{Ballerini2012} found that starspots can affect LD coefficients by up to 30\% in the ultraviolet, with a weaker effect expected at redder wavelengths. If we assume a 10\% variation in the LD coefficients for our $r$-band data, the coefficients move into 1-$\sigma$ agreement between the datasets. 

% \begin{table} \centering
% \caption{\label{tab:search} The parameter search space used in modeling the WASP-19 datasets.}
% \begin{tabular}{lccc} \hline
% Parameter & Lower Boundry & Upper Boundry  \\
% \hline
% Planet/star radii ratio ($r_p/r_s$) & 0.05 & 0.30    \\
% Star + Planet radii ($r_s + r_p$)   & 0.10 & 0.50    \\
% Linear LD coefficient ($u1$) & 0.0 & 1.0  \\
% Quadratic Limb Darkerning Coefficient ($u2$) & 0.0 & 1.0  \\
% Inclination ($i$) in degrees & 70.0 & 90.0  \\
% Transit midpoint offset in phase & -0.5 & 0.5 \\
% Longitude of Spot ($\theta$) in degrees & -90.0 & 90.0 \\
% Latitude of Spot ($\phi$) in degrees & 0.0 (pole) & 90.0 (Equator) \\
% Angular radius of Spot ($r_{\rm spot}$) in degrees & 0.0 & 30.0  \\
% Contrast of Spot ($\rho_{\rm spot}$)  & 0.0 & 1.0  \\
% \hline \end{tabular} \end{table}

\begin{table*} \centering
\setlength{\tabcolsep}{4pt}
\caption{\label{tab:results} Derived photometric parameters from each lightcurve, plus the interval within which the best fit was searched for using {\sc gemc}.}
\begin{tabular}{lccccc} \hline
Parameter & Symbol & Search interval & 2010/02/24 & 2010/02/25  & 2010/02/28  \\
\hline
Radius ratio                  & $r_p/r_s$         & 0.05 to 0.30      &        0.1435 $\pm$ 0.0014  &        0.1417 $\pm$ 0.0013  &        0.1430 $\pm$ 0.0008  \\
Sum of fractional radii       & $r_s + r_p$       & 0.10 to 0.50      &        0.3298 $\pm$ 0.0041  &        0.3300 $\pm$ 0.0025  &        0.3311 $\pm$ 0.0044  \\
Linear LD coefficient         & $u_1$             & 0.0 to 1.0        &         0.314 $\pm$ 0.095   &        0.501  $\pm$ 0.083   &         0.438 $\pm$ 0.077   \\
Quadratic LD coefficient      & $u_2$             & 0.0 to 1.0        &         0.192 $\pm$ 0.023   &        0.222  $\pm$ 0.019   &         0.226 $\pm$ 0.009   \\
Inclination (degrees)         & $i$               & 70.0 to 90.0      &         78.97 $\pm$ 0.39    &        78.92  $\pm$ 0.37    &         78.91 $\pm$ 0.44    \\
Transit epoch (HJD/UTC)       & $T_0$             & $\pm$0.5 in phase & 2455251.79628 $\pm$ 0.00014 & 2455252.58506 $\pm$ 0.00010 & 2455255.74045 $\pm$ 0.00012 \\
Longitude of spot (degrees)   & $\theta$          & -90 to +90        &         -9.54 $\pm$ 0.15    &         14.98 $\pm$ 0.13    &                             \\
Latitude of Spot (degrees)    & $\phi$            & 0.0 to 90.0       &         64.93 $\pm$ 0.32    &         65.37 $\pm$ 0.21    &                             \\
Spot angular radius (degrees) & $r_{\rm spot}$    & 0.0 to 30.0       &         15.01 $\pm$ 0.21    &         15.18 $\pm$ 0.15    &                             \\
Spot contrast                 & $\rho_{\rm spot}$ & 0.0 to 1.0        &         0.777 $\pm$ 0.011   &         0.760 $\pm$ 0.017   &                             \\
\hline \end{tabular} \end{table*}

\subsection{Photometric results}

\begin{figure} \includegraphics[width=0.46\textwidth,angle=0]{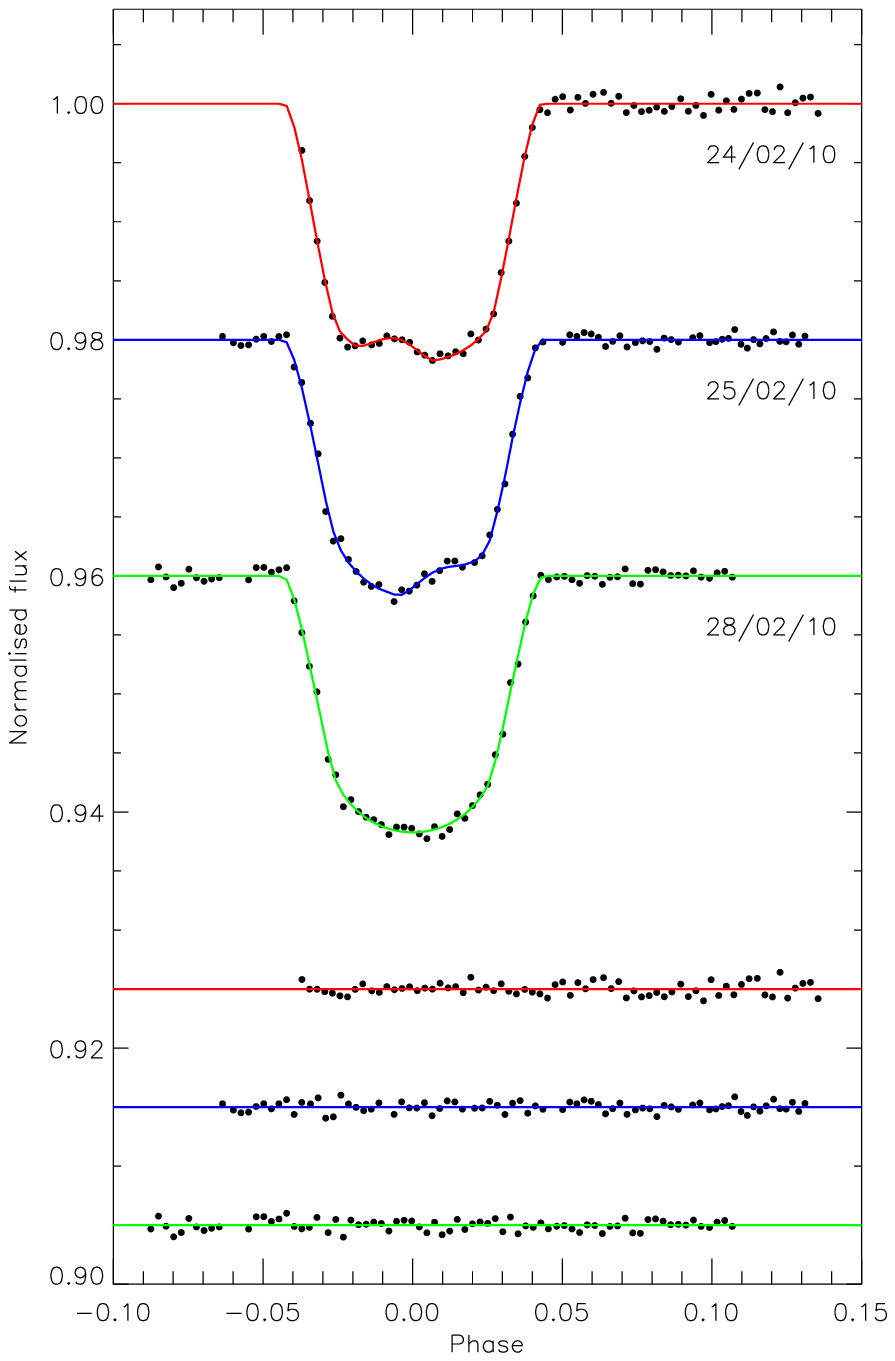} 
\caption{\label{fig:lightcurves} Transit light curves and the best-fitting 
models. The residuls are displayed at the base of the figure.} \end{figure}

As the combined fit to the three datasets has significantly larger residuals than individual fits, we based our final results on the individual fits to the data. The final photometric parameters for the WASP-19 system are given in Table\,\ref{tab:results2} and are weighted means plus 1-$\sigma$ uncertainties of the results from the three individual fits. Fig.\,\ref{fig:lightcurves} compares the light curves to the best-fitting models, including the residuals.

\begin{figure} \includegraphics[width=0.24\textwidth,angle=0]{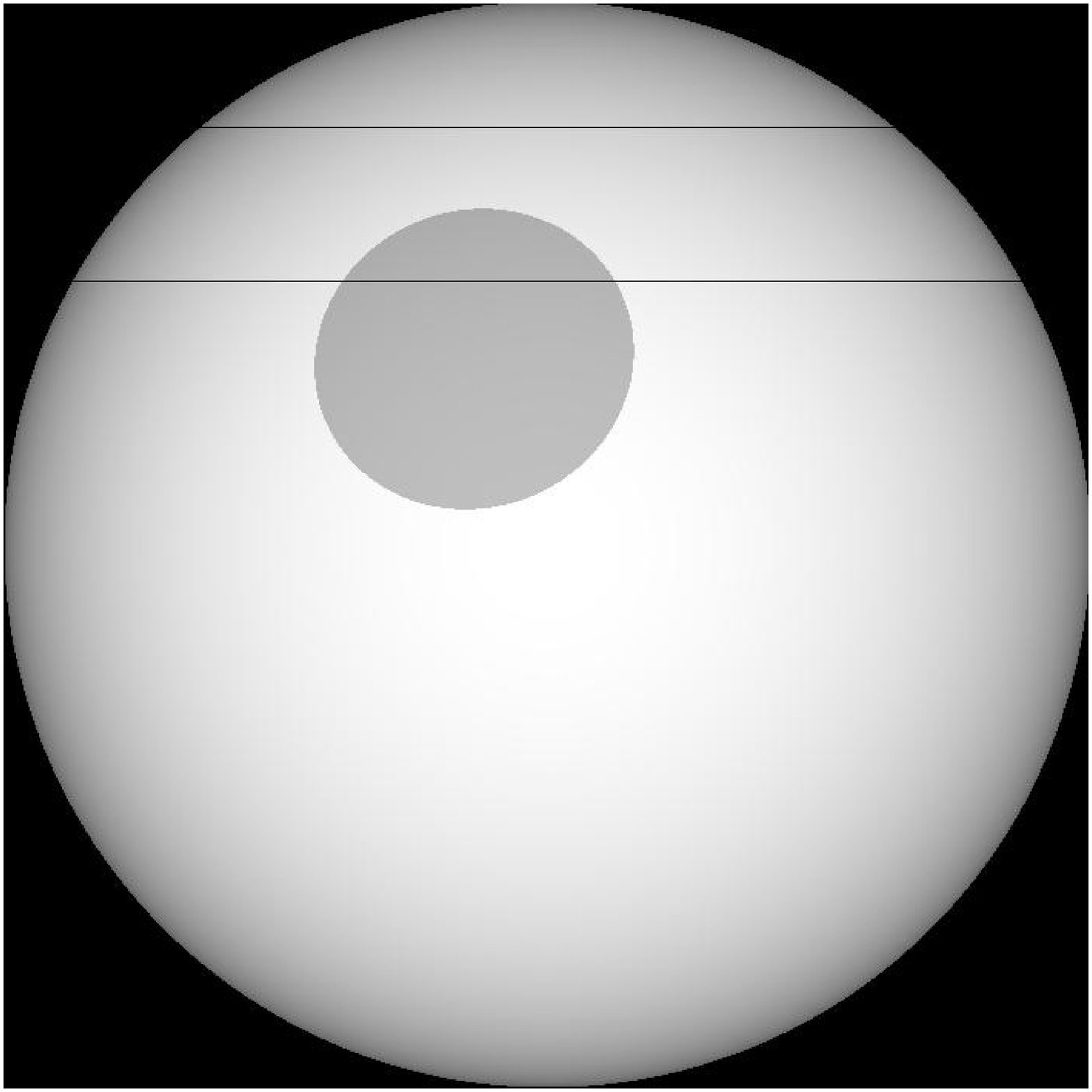} 
               \includegraphics[width=0.24\textwidth,angle=0]{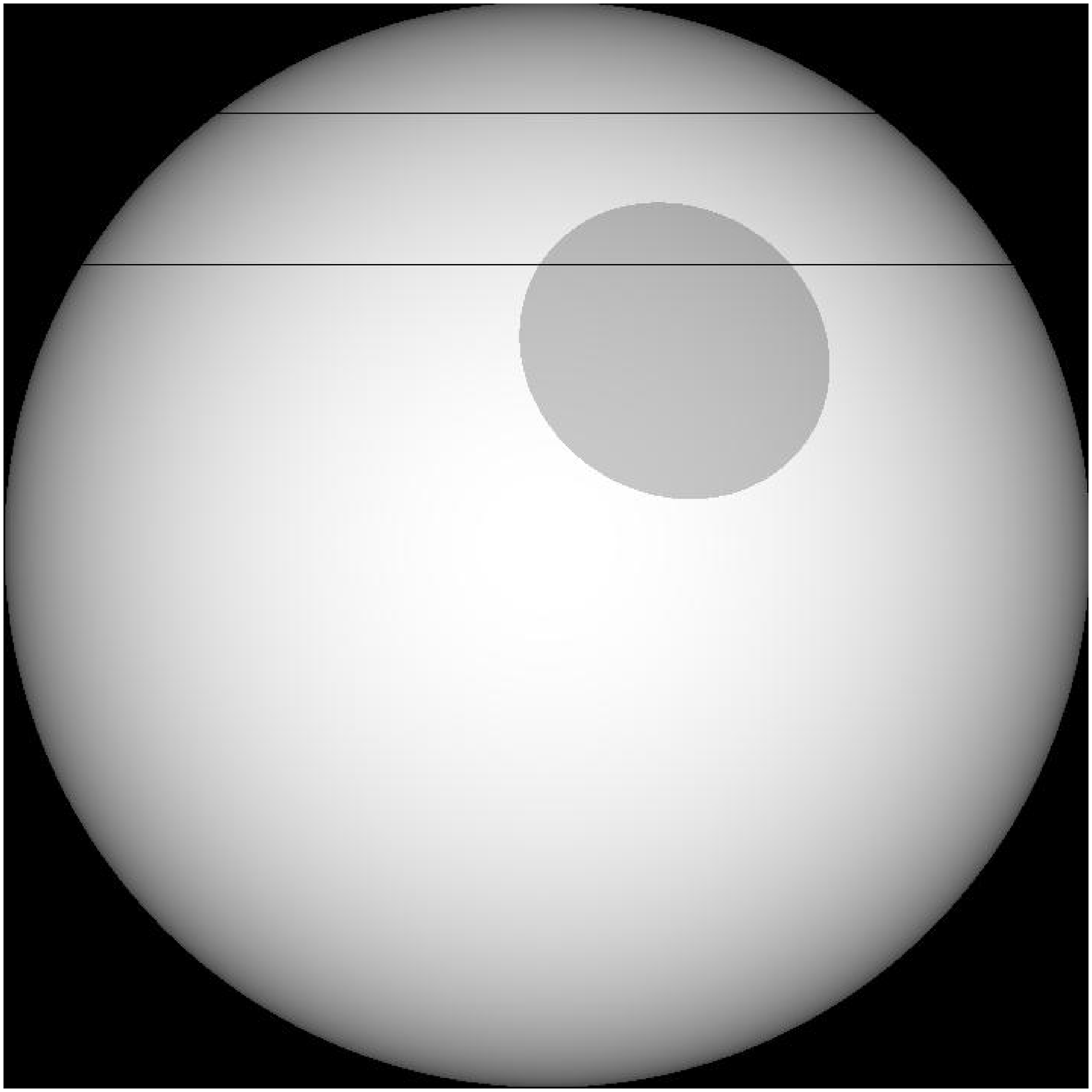}
\caption{\label{fig:spot1} Representation of the stellar disc, starspot and 
transit chord for the two datasets containing spot anomalies.} \end{figure}

The results from modelling the spot anomalies on the nights of 2010/02/24 and 2010/02/25 confirm that they are due to the same spot rotating around the surface of the star, as the spot sizes and contrasts are in good agreement. Fig.\,\ref{fig:spot1} is a representation of the stellar disc, the spot and the transit chord for the two nights of observations. 

From the positions of the starspot at the time of the transits on the nights of 2010/02/24 and 2010/02/25, it is possible to calculate the rotational period of the star and the sky-projected spin orbit alignment of the system using simple geometry. The spot has travelled $24.52^{\circ} \pm 0.28^{\circ}$ in $1.015\pm0.001$ orbital periods, giving a rotational period of $P_{\rm rot} = 11.76\pm0.09$\,d at a latitude of $65^{\circ}$. Combining this with the stellar radius found below, we calculate the latitudinal rotational velocity of the star to be $v_{\left(65^\circ\right)} = 3.88\pm0.15$\kms. The positions of the spot finally yield a sky-projected spin orbit alignment of $\lambda = 1.0^{\circ} \pm 1.2^{\circ}$ for WASP-19.

\begin{figure*} \includegraphics[width=\textwidth,angle=0]{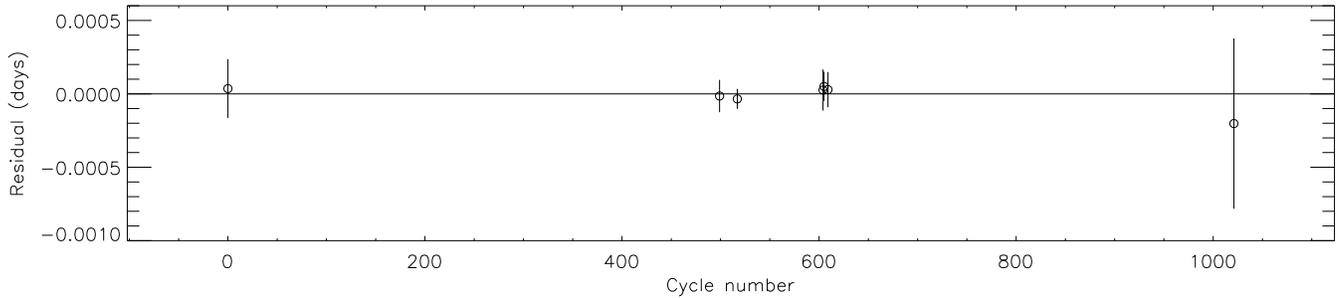} 
\caption{\label{fig:oc} Residuals of the available times of mid-transit 
versus the orbital ephemeris found in this work. The three timings from 
this work are the cluster of three points around cycle number 600.} \end{figure*}

We have collected the available times of mid-transit for WASP-19 from the literature \citep{Hebb2010,Anderson2011,Dragomir2011,Albrecht2012}. All timings were converted to the HJD/TDB timescale and used to obtain a new orbital ephemeris:
$$ T_0 = {\rm HJD/TDB} \,\, 2\,454\,775.33754 (18) \, + \, 0.78883942 (33) \times E $$
where $E$ represents the cycle count with respect to the reference epoch and the bracketed quantities represent the uncertainty in the final digit of the preceding number. Fig.\,\ref{fig:oc} and Table\,\ref{tab:minima} show the residuals of these times against the ephemeris. We find no evidence for transit timing variations in the system.

\begin{table} \centering
\caption{\label{tab:results2} Combined system and spot parameters.}
\setlength{\tabcolsep}{4pt}
\begin{tabular}{lcc} \hline
Parameter & Symbol & Value   \\
\hline
Radius ratio                             & $r_p/r_s$         & 0.1428 $\pm$ 0.0006 \\
Sum of fractional radii                  & $r_s + r_p$       & 0.3301 $\pm$ 0.0019 \\
Linear LD coefficient                    & $u_1$             &  0.427 $\pm$ 0.049  \\
Quadratic LD coefficient                 & $u_2$             &  0.222 $\pm$ 0.008  \\
Inclination (degrees)                    & $i$               &  78.94 $\pm$ 0.23   \\
Spot angular radius (degrees)            & $r_{\rm spot}$    &  15.13 $\pm$ 0.12   \\
Spot contrast                            & $\rho_{\rm spot}$ &  0.771 $\pm$ 0.010  \\
Stellar rotation period (d)              & $P_{\rm rot}$     &  11.76 $\pm$ 0.09   \\
Projected spin orbit alignment (degrees) & $\lambda$         &    1.0 $\pm$ 1.2    \\
\hline \end{tabular} \end{table}

\begin{table} \begin{center}
\caption{\label{tab:minima} Times of minimum light of WASP-19
and their residuals versus the ephemeris derived in this work.
\newline {\bf References:}
(1) \citet{Hebb2010}; 
(2) \citet{Albrecht2012};
(3) \citet{Anderson2011};
(4) This work;
(5) \citet{Dragomir2011}.}
\begin{tabular}{l@{\,$\pm$\,}l r r l} \hline
\multicolumn{2}{l}{Time of minimum}   & Cycle  & Residual & Reference \\
\multicolumn{2}{l}{(HJD/TDB $-$ 2400000)} & no.    & (HJD)    &           \\
\hline
54775.33757 & 0.00020 &     0.0 &  0.00004 & 1 \\   %  0.2 sigma     2010ApJ...708..224Hebb         %
55168.96839 & 0.00011 &   499.0 & -0.00001 & 2 \\   % -0.1 sigma     2012arXiv:1206.6105v2Albrecht  %
55183.16748 & 0.00007 &   517.0 & -0.00003 & 3 \\   % -0.5 sigma     2011arXiv:1112.5145v1Anderson  %
55251.79657 & 0.00014 &   604.0 &  0.00003 & 4 \\   %  0.2 sigma     This                           %
55252.58544 & 0.00010 &   605.0 &  0.00005 & 4 \\   %  0.5 sigma     This                           %
55255.74077 & 0.00012 &   609.0 &  0.00003 & 4 \\   %  0.2 sigma     This                           %
55580.74238 & 0.00058 &  1021.0 & -0.00020 & 5 \\   % -0.3 sigma     2011ApJ...142..115Dragomir     %
\hline \end{tabular} \end{center} \end{table}

\subsection{Physical properties of the WASP-19 system}

\begin{table} \centering
\caption{\label{tab:model} Physical properties of the WASP-19 system.}
\begin{tabular}{l r@{\,$\pm$\,}c@{\,$\pm$\,}l}
\hline 
Parameter & \mcc{Value} \\
\hline
Stellar mass (\Msun)          & 0.904    & 0.040    & 0.021       \\
Stellar radius (\Rsun)        & 1.004    & 0.016    & 0.008       \\
Stellar surface gravity (cgs) & 4.391    & 0.008    & 0.003       \\
Stellar density (\psun)       & \mcc{$0.893 \pm 0.015$}           \\
Planet mass (\Mjup)           & 1.114    & 0.036    & 0.017       \\
Planet radius (\Rjup)         & 1.395    & 0.023    & 0.011       \\
Planet surface gravity (\mss) & \mcc{$14.19 \pm  0.26$}           \\
Planet density (\pjup)        & 0.384    & 0.011    & 0.003       \\
Equilibrium temperature       & \mcc{$2067 \pm   23$}             \\
Safronov number               & 0.02852  & 0.00057  & 0.00023     \\
Semimajor axis (AU)           & 0.01616  & 0.00024  & 0.00013     \\
Age              (Gyr)        & \ermcc{11.5}{ 2.7}{ 2.3}{ 0.7}{ 1.5} \\
\hline \end{tabular} \end{table}

Now we have measured the photometric properties of WASP-19 we can proceed to the determination of its physical characteristics. We undertook this analysis following the method of \citet{Me09mn}, which uses the parameters measured from the light curves and spectra, plus tabulated predictions of theoretical models. We adopted the values of $i$, $r_p/r_s$ and $r_s + r_p$ from Table\,\ref{tab:results2}, and the stellar properties of effective temperature $\Teff = 5440 \pm 60$\,K \citep{Maxted++11mn}, velocity amplitude $K_{\rm s} = 257 \pm 3$\ms\ \citep{Hellier2011} and metal abundance $\FeH = 0.02 \pm 0.09$ \citep{Hellier2011}.

An initial value of the velocity amplitude of the planet, $K_{\rm p}$, was used to calculate the physical properties of the system using standard formulae and the physical constants listed by \citet{Me11mn}. The mass and \FeH\ of the star were then used to obtain the expected \Teff\ and radius, by interpolation within a set of tabulated predictions from stellar theoretical models. $K_{\rm p}$ was iteratively refined until the best agreement was found between the observed and expected \Teff, and the measured $r_{\rm s}$ and expected $\frac{R_{\rm s}}{a}$. This was performed for ages ranging from the zero-age to the terminal-age main sequence, in steps of 0.01\,Gyr. The overall best fit was found, yielding estimates of the system parameters and the evolutionary age of the star.

This procedure was performed separately using five different sets of stellar theoretical models \citep[see][]{Me10mn}, and the spread of values for each output parameter was used to assign a systematic error. Statistical errors were propagated using a perturbation algorithm. An alternative set of physical properties was calculated using a calibration of stellar properties based on well-studied eclipsing binary star systems \citep{Enoch+10aa}, with calibration coefficients from \citet{Me11mn}.

The final results of this process are in reasonable agreement with themselves and with published results for WASP-19. The final physical properties are given in Table.\,\ref{tab:model} and incorporate separate statistical and systematic errorbars for those parameters which depend on the theoretical models. The final statistical errorbar for each parameter is the largest of the individual ones from the solutions using each of the five different stellar models. The systematic errorbar is the largest difference between the mean and the individual values of the parameter from the five solutions. One point to note is that the inferred age of the star is rather large, particularly given its rotation period and activity level. The age is governed primarily by the input \Teff\ and \FeH, so a check of these spectroscopic parameters would be useful.

%%%%%%%%%%%%%%%%%%%%%%%%%%%%%%%%%%%%%%%%%%%%%%%%%%%%%%%%%%%%%%%%%%%%%%%%%%%%%%%%%%%%%%%%%%%%%%%%%%%%%%%%%%%%%%%%%

\section{Summary and discussion}
\label{sec:Conclusions}

We have introduced the \textsc{prism} code to model a planetary transit over a spotted star, and the optimisation algorithm \textsc{gemc} for finding the global best fit and associated uncertainties. While \textsc{gemc} still requires significant computing time to calculate parameter uncertainties via Markov chains, the speed at which it can find the optimal solution is a large improvement over the long burn-in currently required by an MCMC routine.

We have applied \textsc{prism} and \textsc{gemc} to three transit light curves of the WASP-19 planetary system. Two of the light curves are of consecutive transits and show anomalies due to the occultation of a starspot by the planet. The measured latitudes and longitudes of the spot during the two transits were used to calculate the rotation period of the star and the sky-projected obliquity of the system. Our model assumes that the spot anomaly can be represented by a circular spot of uniform brightness. It is quite likely that the ``spot'' is in fact a group of smaller spots with lower contrasts, but investigation of this puts extreme demands on data quality and quantity which are practially impossible to satisfy for ground-based observations.

We find a rotation period of $P_{\rm rot} = 11.76 \pm 0.09$\,d at a latitude of $65^\circ$, whereas \citet{Hebb2010} found a $P_{\rm rot}$ of $10.5 \pm 0.2$\,d from rotational modulation of the star's brightness over several years. The latter value comes from the spot activity over the whole visible surface of the star, whereas our value is for a specific latitude. The difference between these two numbers may therefore indicate differential rotation. \citet{Anderson2011} used the measured Ca\,H\&K line activity index, $\log R^\prime_{\rm HK}$, to infer $P_{\rm rot} = 12.3 \pm 1.5$\,d using the activity--rotation calibration by \citet{Mamajek2008}, which is in good agreement with the values measured by ourselves and by \citet{Hebb2010}.

We find a rotational velocity of $v_{\left(65^\circ\right)} = 3.88 \pm 0.15$\kms\ for WASP-19\,A, which in the absence of differential rotation would yield an equatorial rotation velocity of $v_{\left(90^\circ\right)} = 4.30 \pm 0.15$\kms. \citet{Hellier2011} reported a spectroscopic measurement for $v \sin I$ of $5.0 \pm 0.3$\kms\ and assumed this value represented the equatorial velocity. They included it as a prior when modelling the Rossiter-McLaughlin effect, finding a final value of $v \sin I = 4.6 \pm 0.3$\kms. This last measurement is appropriate for the latitude at which the planet transits, and may differ from ours due to the effect of starspots on radial velocity measurements taken during transit.

We find a sky-projected obliquity of $\lambda = 1.0^{\circ} \pm 1.2^{\circ}$ for WASP-19, which is in agreement with but more precise than published values based on observations of the Rossiter-McLaughlin effect ($4.6^{\circ} \pm 5.2^{\circ}$, \citealt{Hellier2011}; $15^{\circ} \pm 11^{\circ}$, \citealt{Albrecht2012}). $\lambda$ gives the lower boundary of the true spin-orbit angle, $\psi$. As stated by \citet{Fabrycky2009}, finding a small value for $\lambda$ can be interpreted in different ways. The spot method could allow us to determine $\psi$, rather than just $\lambda$, given light curves of three or more transits all showing anomalies due to the same spot. But with only two light curves it is difficult to be sure that $\psi$ lies close to $\lambda$. We calculated a minimum rotation period of WASP-19 of 5.5\,d, for the extreme case that the orbital axis was aligned with the line of sight. This result disagrees with previous measurements \citep{Hebb2010,Anderson2011}. Whilst we are unable to determine the true value of $\psi$ with the data in hand, we find no evidence for a spin-orbit misalignment in the WASP-19 system. With a low obliquity and cool host star, WASP-19 follows the idea put forward by \citet{Winn2010c} that planetary systems with cool stars will have a low obliquity. It also lends weight to the idea that WASP-19\,b formed at a much greater distance from host star and suffered orbital decay through tidal interactions with the protoplanetary disc.

%%%%%%%%%%%%%%%%%%%%%%%%%%%%%%%%%%%%%%%%%%%%%%%%%%%%%%%%%%%%%%%%%%%%%%%%%%%%%%%%%%%%%%%%%%%%%%%%%%%%%%%%%%%%%%%%%%%%%%%%%%%%%%%%%%%%%%%%%%%%%%%%%%%%%

\section{Acknowledgements}

Based on observations made with ESO Telescopes at the La Silla Observatory under programme ID 084.D-0056(A). JTR acknowledges financial support from STFC in the form of an Ph.D. Studentship.

%%%%%%%%%%%%%%%%%%%%%%%%%%%%%%%%%%%%%%%%%%%%%%%%%%%%%%%%%%%%%%%%%%%%%%%%%%%%%%%%%%%%%%%%%%%%%%%%%%%%%%%%%%%%%%%%%%%%%%%%%%%%%%%%%%%%%%%%%%%%%%%%%%%%%

%\bibliographystyle{/home/jtr/plainnatjtr}
%\bibliographystyle{mn_new}
%\bibliography{aamnem99,jtr}
%\bibliography{iau_journals,Lit}
%\bsp

%%%%%%%%%%%%%%%%%%%%%%%%%%%%%%%%%%%%%%%%%%%%%%%%%%%%%%%%%%%%%%%%%%%%%%%%%%%%%%%%%%%%%%%%%%%%%%%%%%%%%%%%%%%%%%%%%%%%%%%
\end{document}